\documentstyle[12pt]{article}
\def\secwidth  {15.3cm} 
\def\lb       {\left( }
\def\rb       {\right) }

\def\lbb     {\left[ }
\def\rbb      {\right] }
\def\comma      { \, , }
\def\period     { \, . }
\def\bra#1      { \langle \, #1 \, \vert \, }
\def\ket#1      { \, \vert \, #1 \, \rangle \, }
\def\semiket#1  { \, #1 \, \rangle \, }

\def\half       {  {1\over 2}  }

\def\abs#1      {  \, \vert #1 \vert \,   }
\def\Im#1    { \, {\rm Im } \, #1  }
\def\Re#1    { \, {\rm Re}  \, #1  }

\def\bfZ     { {\bf Z}}

\def\vecii#1#2      {  \left(\begin{array}{c}#1\\#2\end{array}\right)  }
\def\veciii#1#2#3   {  \left(\begin{array}{c}#1\\#2\\#3\end{array}\right)  }
\def\matrixii#1#2#3#4            {  \left(\begin{array}{cc}#1&#2\\#3&#4
                                       \end{array}\right) }
\def\matrixiii#1#2#3#4#5#6#7#8#9 {  \left(\begin{array}{ccc}#1&#2&#3\\
                                     #4&#5&#6\\#7&#8&#9\end{array}\right)  }
\def\eqabegin         {  \begin{eqnarray}  }
\def\eqaend           {  \end{eqnarray}  }
\def\nn               {  \nonumber  }

\def\sectionnumbering { \setcounter{equation}{0}
         \renewcommand{\theequation}{\arabic{section}.\arabic{equation}}}

\def\mysection#1{\addtocounter{section}{1} \setcounter{subsection}{0}
                 \sectionnumbering 
    {\large\bf \arabic{section} \quad  
     \parbox[t]{\secwidth}{#1}  }   \par \noindent}

\def\csectionast#1    { \begin{center}  
  \noindent  {\large\bf #1  }   \par \noindent \end{center} }
\renewcommand{\thefootnote}{\fnsymbol{footnote}}

\def\xxx#1 {{\tt hep-th/#1}}
\def\grqc#1 {{\tt gr-qc/#1}}
\def\npb#1(#2)#3 { Nucl. Phys. {\bf B#1} (#2) #3 }
\def\rep#1(#2)#3 { Phys. Rept.{\bf #1} (#2) #3 }
\def\plb#1(#2)#3{Phys. Lett. {\bf #1B} (#2) #3}
\def\prl#1(#2)#3{Phys. Rev. Lett.{\bf #1} (#2) #3}
\def\prd#1(#2)#3{Phys. Rev. {\bf D#1} (#2) #3}
\def\ap#1(#2)#3{Ann. Phys. {\bf #1} (#2) #3}
\def\rmp#1(#2)#3{Rev. Mod. Phys. {\bf #1} (#2) #3}
\def\cmp#1(#2)#3{Comm. Math. Phys. {\bf #1} (#2) #3}
\def\mpl#1(#2)#3{Mod. Phys. Lett. {\bf A#1} (#2) #3}
\def\ijmp#1(#2)#3{Int. J. Mod. Phys. {\bf A#1} (#2) #3}
\def\mpla#1(#2)#3{Mod. Phys. Lett. {\bf A#1} (#2) #3}
\def\jhep#1(#2)#3{JHEP {\bf  #1} (#2) #3}
\def\cqg#1(#2)#3{Class. Quant. Grav {\bf  #1} (#2) #3}
\def\kket#1   { \vert #1 \rangle \! \rangle }
\def\bbra#1   { \langle \! \! \langle  #1 \vert }
\def\qtilde   { \tilde{q} }
\def\Dlw      { {\cal D}_{\rm lw} }
\def\Dhw      { {\cal D}_{\rm hw} }
\begin{document}
%
\def\papertitlepage{\baselineskip 3.5ex \thispagestyle{empty}}
\def\preprinumber#1#2#3#4{\hfill \begin{minipage}{4.2cm}  #1
              \par\noindent #2
              \par\noindent #3
              \par\noindent #4
             \end{minipage}}
\renewcommand{\thefootnote}{\fnsymbol{footnote}}
%
%
\papertitlepage
\setcounter{page}{0}
\preprinumber{January 2000}{UTMS 2000-4}{UTHEP-416}{hep-th/0001063}
\baselineskip 0.8cm 
\vspace{2.0cm}
\begin{center}
{\large\bf Modular invariance of string theory on $AdS_3$}
\end{center}
\vskip 7ex
\baselineskip 1.0cm
\begin{center}
     {\sl Akishi ~Kato}
   \footnote[2]{akishi@ms.u-tokyo.ac.jp} \ \ {and} \ \
     {\sl Yuji  ~Satoh}
   \footnote[3]{ysatoh@het.ph.tsukuba.ac.jp}  \\
 \vskip 3ex
    {\it $^{\dagger}$Graduate School of Mathematical Science, 
   University of Tokyo} \\
  \vskip -2ex
         {\it Komaba, Meguro-ku, Tokyo 153-8914, Japan} \\
\vskip 1ex
    {\it $^{\ddagger}$Institute of Physics, University of Tsukuba} \\
 \vskip -2ex
   {\it Tsukuba, Ibaraki 305-8571, Japan} \\
\end{center}
\vskip 8ex
%
\baselineskip=3.5ex
\begin{center} {\large\bf Abstract} \end{center} 
\vskip 2ex
 
We discuss the modular invariance of the $SL(2,R)$ WZW model.
In particular, we discuss in detail the modular invariants 
using the $\hat{sl}(2,R) $ characters based on  the discrete 
unitary series of the $SL(2,R)$ representations. The explicit forms 
of the corresponding characters are known when no singular vectors 
appear. We show, for example,  that from such characters 
modular invariants can be obtained only when the level 
$k<2$ and infinitely large spins are included. In fact, we give a modular 
invariant with three variables $Z(z,\tau, u)$ in this case.
We also argue that the discrete series characters are not sufficient 
to construct a modular invariant compatible with the unitarity bound, 
which was proposed to resolve the ghost problem of the $SL(2,R)$ strings.
%
%
%
%
%
%
\newpage
\renewcommand{\thefootnote}{\arabic{footnote}}
\setcounter{footnote}{0}
\setcounter{section}{0}
\baselineskip = 0.6cm
\pagestyle{plain}
\noindent
\mysection{Introduction}
The string theory on $AdS_3$, namely, $SL(2,R)$ is important in various 
respects and it has been investigated for more than a decade 
\cite{BFOW}-\cite{Teschner1}.
It gives us the simplest string model in 
backgrounds with curved time. Without R-R charges, the system is 
described by the $SL(2,R)$ WZW model. This WZW model is one of the simplest 
models of the non-compact CFT. In addition, it is closely related to 
the string theory in some two ($SL(2,R)/U(1)$) and three (BTZ) dimensional 
black hole backgrounds. The appearance of the AdS/CFT correspondence 
\cite{AdSCFT} aroused the renewed interest in the $SL(2,R)$ WZW model and
its Euclidean version, the $SL(2,C)/SU(2)$ WZW model, e.g., 
\cite{GKS}-\cite{GN}.  

However, in spite of recent intensive studies, there still remain 
open questions for these string models at the fundamental level.
Such a state of the problem was discussed in \cite{PMP}.
In fact, soon after the study of the $SL(2,R)$ WZW model was initiated,
it was found that the model contains negative-norm physical states (ghosts)
\cite{BFOW}. So far, there are two types of the proposals for the resolution.
In one proposal \cite{ng1}, the discrete unitary series of the 
$SL(2,R)$ representations is used and it is claimed that 
ghosts can be removed  if the spectrum is truncated so that 
the spin $j$ and the level $k$ of the current algebra 
$\hat{sl}(2,R)$ satisfy
\eqabegin
  && \half \leq -j < \frac{k}{2} \comma \quad k > 2  \label{ub} 
  \period
\eqaend 
(For details of $\hat{sl}(2,R)$ and our conventions, 
see the next section.) The condition (\ref{ub}) is called the unitarity bound.
In the other proposal \cite{Bars,YS}, 
the principal continuous series and the free field
representations of the current algebra are used. 
However, in both cases, it is still an open question if such 
proposals are compatible with other consistency conditions of 
string theory such as the modular invariance and the closure of
the operator product expansion. This is because such consistency 
conditions are not well understood either. Thus it seems 
important to accumulate precise knowledge about them.

In the literature, there were several arguments about this issue.
Regarding the OPE, see \cite{Teschner1}\cite{Teschner2}-\cite{GN}.
On the modular invariance, for instance, 
the modular invariants were constructed 
for the so-called admissible representations in \cite{KW,KS}. 
In \cite{HHRS}, it was argued that modular invariants can be obtained 
for the discrete unitary series by adding the sectors corresponding 
to some winding modes.
In \cite{Huitu}, the modular invariants of the $N=2$ SCFT 
were expressed in terms of the $\hat{sl}(2,R)$ characters based on 
the relationship  between the $N=2$ superconformal algebra and 
$\hat{sl}(2,R)$ \cite{DLP}. In \cite{YS}, modular invariants
for the principal continuous series were discussed along the line 
of \cite{Bars}. 

In this paper, we discuss in detail the possibility of constructing 
modular invariants using the discrete
series characters, which was also discussed in \cite{PMP}. We will
not include additional sectors as in \cite{HHRS}. In the next section,
we review the basics of $SL(2,R)$ and $\hat{sl}(2,R)$ and give 
the modular transformations of the discrete series characters when 
no singular vectors appear.
In section 3, we discuss the case in which finite number of the characters
are used.
In section 4, we discuss the case in which infinite number of the 
characters without singular vectors are used. In section 5, we give 
a summary of our results and discuss the implication to the unitarity 
bound (\ref{ub}).    
\par\vskip 6ex\noindent
\mysection{Setup}
The $SL(2,R)$ current algebra is defined by the commutation relations
\eqabegin
 \lbb J^0_n , J^0_m \rbb & = & - \half  k n \delta_{n+m} \comma  \qquad
 \lbb J^0_n , J^\pm_m \rbb \ = \  \pm J^\pm_{n+m} \comma  \nn \\
 \lbb J^+_n , J^-_m \rbb & = & - 2 J^0_{n+m} + k n \delta_{n+m}
     \period 
\eqaend
The generators of the associated Virasoro algebra are given by
\eqabegin
  L_n & = & \frac{1}{k-2} \sum_{m \in \bfZ } 
   : \half J^+_{n-m} J^-_{m} + \half J^-_{n-m} J^+_{m}
     - J^0_{n-m} J^0_{m}: 
   \comma 
\eqaend
and its  central charge is
\eqabegin
  c &=& \frac{3k}{k-2} \period
\eqaend

The current algebra $\hat{sl}(2,R)$ contains zero mode subalgebra
$SL(2,R)$ generated by $J^{0}_{0}$ and $J^{\pm}_{0}$. In particular, 
for a given $\hat{sl}(2,R)$ representation $V$, its ground state
subspace $V^{0} \equiv \{v \in V | L_{0}v = \Delta_{j} v\}$ with
$\Delta_{j}=-\frac{j(j+1)}{k-2}$ naturally provides a representation of
$SL(2,R)$. The operators $\vec{J}^2$ and $J^{0}_{0}$ act on the states
in $V^{0}$ as
\eqabegin
   \vec{J}^2 \ket{ j,m } = -j(j+1) \ket{ j,m } \comma &&
   J^0_0 \ket{ j,m } = m \ket{ j,m } \period
\eqaend
The other states in $V$ are obtained by
acting on $ \ket{ j,m } \in V^{0}$ with $J^a_{-n} $ $ (n \geq 0)$.

On physical grounds, we are interested in the $\hat{sl}(2,R)$ 
representations $V$ where $V^{0}$ are irreducible and unitary 
as $SL(2,R)$ modules.
In such cases, since $-j(j+1)$ is real and invariant under $ j \to -j-1 $,
one can assume either $j \leq -1/2$ or $j=-1/2+ i\lambda, \lambda\geq 0$
without loss of generality.  

For the universal covering group of $SL(2,R)$, 
there are five classes of such representations: 
\par\medskip\noindent
\hspace*{0.4em} 
(1) \  Identity representation: the trivial representation 
       with $\vec{J}^2 = J^0_0 = 0$.
\par\noindent 
\hspace*{0.4em} 
 (2) \ Principal continuous series: representations 
     with $ m = m_0 + n, \, 0 \leq m_0 < 1, \, n \in \bfZ $ 
   \\ \hspace*{2.4em}  
     and $ j = -1/2 + i \lambda, \, \lambda > 0 $.
\par\noindent
\hspace*{0.4em} 
 (3) \  Supplementary series: representations  
     with $ m = m_0 + n, \, 0 \leq m_0 < 1, \, n \in \bfZ $ \, and 
  \\ \hspace*{2.4em}
   min$\{ -m_0, m_0-1 \} < j \leq -1/2 $.
\par\noindent
\hspace*{0.4em} 
 (4) \  Highest weight discrete series ($\Dhw $): representations 
      with $ m = M_{\rm max} - n, \,n = 0,1,2,...,
 \\ \hspace*{2.4em}
   j = M_{\rm max} \leq -1/2$ and the highest weight state satisfying 
   $J^+_0 \ket{ j,j } = 0$.
\par\noindent
\hspace*{0.4em} 
 (5) \ Lowest weight discrete series ($\Dlw $) representations 
    with $ m = M_{\rm min} + n, \, n = 0,1,2,...,
   \\ \hspace*{2.4em}
   j = - M_{\rm min} \leq -1/2$ and the lowest weight state satisfying 
  $J^-_0 \ket{ j,-j } = 0$.

\medskip \noindent
If we do not take the universal covering group, the parameters are restricted 
to $m_0 = 0, 1/2$ in (2),
$ m_0 = 0 $ in (3) and $ j = $ (half integers) in (4) and (5). 

In the following we will focus on the $\hat{sl}(2,R)$ representations
whose ground state subspace corresponds to the discrete series of the type
($\Dhw $) or ($ \Dlw $); they will be denoted by $V_{j}^{\rm hw}$ or
$V_{j}^{\rm lw}$, respectively.

For, e.g., the $SU(2)$ current algebra, the characters are naturally defined
using three variables (see, for example, \cite{Kac}). 
With this in mind, we define the irreducible characters by
\eqabegin
  {\rm ch}_j (z,\tau,u) & \equiv & e^{2\pi i k u} \;\sum_{V_{j}}\;
e^{-2 \pi i J^0_0 z}
   e^{2 \pi i \tau (L_0 - \frac{c}{24})}
  \period \label{chj}
\eqaend
where the summation is taken over $V_{j}$ which is  
the irreducible representation 
of the type of either $V_{j}^{\rm hw}$ or $V_{j}^{\rm lw}$.
The plus sign in the first factor $ e^{+2 \pi i k u } $
is due to the change $ k \to - k $ compared with the compact case.
Since the $SL(2,R)$ current module has infinite degeneracy with respect 
to $L_0$, it is inevitable to keep $z \neq 0$ for the convergence
of the charcaters. In addition, it turn out that, as functions of the only two 
variables $(\tau,z)$, one cannot obtain proper modular transformations
of the characters. Thus the use of the three variables is essential.

To calculate these characters, one needs to know about singular vectors.
It is known that the $\hat{sl}(2,R)$ 
Verma module at level $k$ with the highest weight 
$|j,j\rangle$ has singular vectors if and only if at least
one of the following conditions is satisfied \cite{singularvec}:
\eqabegin
   (1) && 2 j + 1 = s + (k-2)(r-1) \comma \nn \\
   (2) && 2 j + 1 =  -s - r (k-2) \comma \label{singvec} \\
   (3) && k - 2 = 0  \comma \nn
\eqaend 
where $r,s$ are positive integers. 
For example, in the case of $k-2>0$ and $j \leq -1/2$, the singular vectors
appear when $ 2 j + 1 = -s - r(k-2)$.

Thus, for generic values of $j$, 
there are no singular vectors in
$V_{j}^{\rm hw}$; the irreducible character ${\rm ch}_{j}$ 
coincides with that of the Verma module: \cite{character,HHRS}
\eqabegin
   \chi^{\rm hw}_\mu (z,\tau,u) &\equiv& e^{2 \pi i k u}
    e^{-2 \pi i \mu  z} q^{-\frac{ \mu^2}{k-2}} \ 
   i \vartheta_1^{-1}(z \vert \tau) 
    \period \label{hwchi}
\eqaend
Here $ q = e^{2 \pi i \tau}$, $\mu = j + 1/2$ and 
\eqabegin 
   \vartheta_1 (z \vert \tau) & = & 2 q^{1/8} \sin (\pi z) 
  \prod_{n=1}^\infty (1-q^n)(1-q^n e^{2\pi i z})(1-q^n e^{-2\pi i z}) 
  \period \label{theta1}
\eqaend
Similarly, for $V_{j}^{\rm lw}$ we have $\chi^{\rm lw}_{\mu} (z,\tau,u) = 
\chi^{\rm hw}_\mu (-z,\tau,u)$.

So far we have considered $\Dhw$ and $\Dlw$ with $\mu\equiv
j+1/2 \leq 0$ separately. However with the help of the symmetry $\chi^{\rm
lw}_{\mu} (z,\tau,u)$ $=$ $\chi^{\rm hw}_{\mu} (-z,\tau,u) = - \chi^{\rm
hw}_{-\mu} (z,\tau,u) $, we extend the range of $\mu$ as
$-\infty<\mu<+\infty$ and drop the superscript
$\rm{hw}$ with the following convention:
\eqabegin 
\chi_{\mu}(z,\tau,u)\equiv\mbox{\ r.h.s.\ of \  (\ref{hwchi}) \ }=
\left\{\begin{array}{ll}
	    +\chi^{\rm hw}_{\mu}(z,\tau,u),& (\mu \leq 0)\\[.5em]
	    -\chi^{\rm lw}_{-\mu}(z,\tau,u).& (\mu \geq 0)
       \end{array}\right.\nn
\nn
\eqaend
We remark that one cannot consider the specialized characters 
$\chi_\mu (0, \tau, 0)$ since they diverge in the limit $z \to 0$
because of the infinite degeneracy with respect to $L_0$.

For a special values of $\mu = j + 1/2$ for which the Verma module has
singular vectors, the irreducible character ${\rm ch}_j$ is different from
$\chi_{\mu}$.  In such a case, the explicit form of ${\rm ch}_j$
 seems unknown except for several cases.  

In our normalization of $(z,\tau,u)$, 
the modular transformations are generated by  
\eqabegin
        S :  && (z, \tau, u ) \quad \to \quad     
                   \Bigl( \, \frac{z}{\tau}, -\frac{1}{\tau}, 
   u + \frac{z^2}{4\tau} \, \Bigr) \comma \nn \\
        T :  && (z, \tau, u ) \quad \to \quad (z, \tau + 1, u) 
    \period
\eqaend
Under $T$-transformation, the characters $\chi_\mu$ just get phases, 
\eqabegin
   \chi_\mu (z, \tau + 1, u) &=& 
   e^{-2 \pi i \lb \frac{\mu^2}{k-2} + \frac{1}{8}\rb }
    \chi_\mu (z, \tau, u)
 \period \label{T}
\eqaend
For $ k-2 < 0 $, the $S$-transformation of $\chi_\mu (z,\tau,0) $
is given in \cite{PMP}. In our case with three variables, it reads
as
\eqabegin
  \chi_\mu \Bigl( \,  \frac{z}{\tau}, -\frac{1}{\tau}, 
   u + \frac{z^2}{4\tau} \, \Bigr) & = &
  \sqrt{\frac{-2}{2-k} } \int_{-\infty}^\infty d\nu 
  \ e^{4 \pi i \frac{\mu \nu}{k-2}} \chi_\nu (z,\tau,u)
  \period \label{S1}
\eqaend
For $k-2 >0 $, the right-hand side of (\ref{S1}) does not make sense
since $\Delta_{j}\to -\infty$ as $\nu\to \pm \infty$ and the 
integral diverges.
Instead, after some calculation, we get a slightly different
formula which does converge:
\eqabegin
  \chi_\mu \Bigl( \, \frac{z}{\tau}, -\frac{1}{\tau}, 
   u + \frac{z^2}{4\tau} \, \Bigr) &=&
  \sqrt{\frac{-2}{k-2} } \int_{-\infty}^\infty d\nu 
  \ e^{- 4 \pi \frac{\mu \nu}{k-2}} \chi_{i\nu} (z,\tau,u)
  \period \label{S2}
\eqaend
Note that an imaginary $\mu = j + 1/2 $ corresponds to a spin 
of the principal continuous series but $\chi_{i\mu}$ are { not}
the characters for those representations any more.
%

\newpage
\mysection{Modular invariants from finite number of discrete series
characters}
\par\noindent
Now let us start the discussion of the modular invariance. 
In this section, we will discuss the possibility of constructing 
the modular invariants using 
finite number of the discrete series characters.

First, we would like to discuss the modular invariants using
the characters in generic cases, $\chi_\mu$.
We then show that  
{\it it is impossible to make modular invariants
from finite number of $\chi_\mu $;
more precisely, a finite sum of the left and right characters
\eqabegin
   Z(z,\tau,u) & \equiv & \sum_{\mu,\mu'} N_{\mu,\mu'} 
   \chi_{\mu} (z,\tau,u) \chi^*_{\mu'} (z,\tau,u)
  \comma \label{Zchi}
\eqaend
with non-zero coefficients $N_{\mu,\mu'}$ cannot be modular invariant.} 
The argument is a simple application of Cardy's  
for $c>1$ CFT \cite{Cardy}. We compute in two different ways
the leading behavior of $Z$ in the 
limit $ \tau \to + i 0 $ with $ \abs{ z/\tau } $ fixed.
On one hand,
using the $S$-transform of $\vartheta_1$, one finds 
\eqabegin
   Z(z,\tau,u) &=& e^{- 4 \pi k \Im{ u } } 
    \ e^{-2\pi \Im{ \frac{z^2}{\tau} } } 
     \abs{ \, \tau \ \vartheta_1^{-2} 
 \bigl( \, \frac{z}{\tau} \vert -\frac{1}{\tau} \, \bigr) \, } 
  \sum N_{\mu,\mu'} \, e^{-2\pi i (\mu z - \mu' z^*)} q^{-\frac{\mu^2}{k-2}} 
    (q^*)^{-\frac{\mu'^2}{k-2}} 
    \comma \nn \\
 & \sim & e^{- 4 \pi k \Im{ u } } \lb \sum N_{\mu,\mu'} \rb
       \abs{ \, \frac{\tau}{4} \sin^{-2} (\pi z/\tau) \, } 
  \ \qtilde ^{-1/4}    
  \comma \label{limit1}
\eqaend
where $ \qtilde = e^{-2\pi i/\tau } $.
On the other hand, if the partition function is modular invariant,
one should get
\eqabegin
  Z(z,\tau,u) &=& Z \bigl( \, \frac{z}{\tau}, -\frac{1}{\tau}, 
   u + \frac{z^2}{4\tau} \, \bigr) \label{limit2} \\
   & \sim & e^{- 4 \pi k \Im{ u } }  \ 
  e^{-2 \pi i ( \mu_0 \frac{z}{\tau} - \mu'_0 (\frac{z}{\tau})^* ) }
   N_{\mu_0,\mu_0'} \frac{1}{4} \abs{ \sin^{-2}(\pi z/\tau) \, }
   \ \qtilde^{-\frac{\mu_0^2 + \mu'^2_0}{k-2} - \frac{1}{4}} 
   \comma \nn 
\eqaend
where the pair $(\mu_0,\mu'_0)$ is 
chosen so that $ (\mu_0^2 + \mu'^2_0)/(k-2)$
takes the maximum value there.
Clearly the two behaviors (\ref{limit1}) and (\ref{limit2}) cannot 
be compatible and hence the statement was shown.  

Next, we would like to consider more general cases including the 
representations with singular vectors. In such cases, the characters
are not always given by $\chi_\mu$ and we need to consider  
the modular invariant 
partition functions in terms of the irreducible characters ${\rm ch}_{j}$ 
rather than $\chi_{\mu}$:
\eqabegin
   \hat{Z}(z,\tau,u) & \equiv & \sum_{\mu,\mu'} N_{\mu,\mu'} 
   {\rm ch}_{j} (z,\tau,u) {\rm ch}^*_{j'} (z,\tau,u) 
  \comma \label{Zch}
\eqaend
where $N_{\mu,\mu'} > 0 $.

We now show
that, {\it for $k > 2$, it is impossible to construct modular invariants
from finite number of the characters for $V_{j}^{\rm hw}$ 
(or $V_{j}^{\rm lw}$) only}. 
We prove only the case of $V_{j}^{\rm hw}$ since the argument 
for $V_{j}^{\rm lw}$ is similar. 
To this end, we decompose the character 
${\rm ch}^{\rm hw}_{j}(z,\tau,u)$ 
in terms of the variable $z$ :
\eqabegin
  {\rm ch}^{\rm hw}_j (z,\tau,u) & = & e^{2\pi i k u} e^{-2 \pi i j z}
   q^{-\frac{\mu^2}{k-2} -\frac{1}{8}} \sum_{p \in \bfZ} e^{2 \pi i p z}
    {\rm ch}_{j,p} (q)
  \period \label{chp}
\eqaend
Note that 
${\rm ch}_{j,p} (q)$ 
is a power series in $q$ with non-negative integer coefficients.
In the generic case when there are no singular vectors, i.e., 
${\rm ch}^{\rm hw}_j = \chi_\mu$, we denote the above expansion by
\eqabegin
  \chi_\mu (z,\tau,u) & = & e^{2\pi i k u} e^{-2 \pi i j z}
   q^{-\frac{\mu^2}{k-2} -\frac{1}{8}} \sum_{p \in \bfZ} e^{2 \pi i p z}
    \chi_p (q)
  \period \label{chip}
\eqaend
Comparing this with (\ref{hwchi}) gives
\eqabegin
  \sum_{p \in \bfZ} e^{2 \pi i p z} \chi_p (q)
  &=& e^{-\pi i z} q^{1/8}i \vartheta_1^{-1}(z \vert \tau)
  \period
\eqaend
From the explicit form of the theta function (\ref{theta1}), 
one finds that the expansion (\ref{chip}) converges absolutely
(at least) for 
\eqabegin
   &&  0 \, < \, \Im{ z } \, < \, \Im{ \tau }  \comma  \quad 
  \Re{ z } \, = \, \Re{ \tau } \, = \, 0 
  \period
  \label{imaginaryztau} 
\eqaend
Since any irreducible character ${\rm ch}_{j}$ is 
obtained by subtracting from $\chi_\mu$ the contribution from 
singular vectors,
we obtain the inequality
\eqabegin
  0 \ < \ {\rm ch}_{j,p} (q) & \leq & \chi_p (q) \comma \label{chchi}
\eqaend
if ${\rm Re~} \tau =0, ~ {\rm Im~} \tau >0$.
Therefore one can also evaluate the expansion (\ref{chp}) in the region
(\ref{imaginaryztau}) and finds that  
\eqabegin
   0 \ < \ e^{-2\pi i k u} {\rm ch}^{\rm hw}_j (z,\tau,u)
  & \leq & e^{-2\pi i k u} \chi_\mu (z,\tau,u)
  \period \label{chjleqchi}
\eqaend
From the inequality (\ref{chjleqchi}), it follows that in the region
(\ref{imaginaryztau})
\eqabegin
    \hat{Z}(z,\tau,u) & \leq & Z (z,\tau,u)
    \period
\eqaend 
If the partition function is modular invariant, i.e., 
$ \hat{Z} (z,\tau,u) = \hat{Z} \bigl( \, \frac{z}{\tau}, -\frac{1}{\tau}, 
   u + \frac{z^2}{4\tau} \, \bigr) $, in the limit
$ \tau \to + i 0 $ with $ \abs{ z/\tau } $ fixed, the above inequality
gives
\eqabegin
     e^{-2 \pi i ( \mu_0  - \mu'_0 ) z/\tau }
      F\big(\frac{z}{\tau}\bigr) N_{\mu_0,\mu'_0} 
    \qtilde^{-\frac{\mu_0^2+\mu'^2_0}{k-2} - \frac{1}{4}}
       & \leq &   \frac{1}{4}  \lb \sum N_{\mu,\mu'} \rb
       \sin^{-2} (\pi z/\tau) 
   \abs{ \tau } \ \qtilde ^{-1/4}
  \comma
  \label{inequality}
\eqaend
where $F(z/\tau)$ is a function of $ e^{2 \pi i z/\tau}$
which remains finite in the limit.
The inequality (\ref{inequality}) cannot be 
satisfied for $ k> 2$ and hence the statement was shown.

The above argument cannot be generalized to 
the case in which the partition function includes
the characters for  both $V_{j}^{\rm hw}$ and $V_{j}^{\rm lw}$.
This is because the series in (\ref{chj}) can be defined  
in $ \Im{ z } \geq 0 $ for $V_{j}^{\rm hw}$ whereas it can be defined 
in $  \Im{ z } \leq 0 $ for $V_{j}^{\rm lw}$ and hence we cannot find the 
region where the partition function becomes real 
such as (\ref{imaginaryztau}). In this case, the partition function 
is not analytic in $z$ unless the analytic continuation is possible.
To further discuss this case, we may need to 
find the explicit expressions of the irreducible characters 
${\rm ch}_j^{\rm hw(lw)}$.  

One might wonder also if a similar statement holds for $ k<2$.
To check this, let us note that the arguments in this section hold
for more generic highest and lowest weight representations besides
$V_{j}^{\rm hw}$ and $V_{j}^{\rm lw}$ since 
we did not use any specific properties of 
the unitary representations $\Dhw$ and $\Dlw$. 
However, for $k<2$, modular invariants 
using finite number of the characters are 
actually known \cite{KW,KS} for the admissible representations.
Thus a simple inequality
(\ref{chchi}) should not exclude the possibility of modular invariants
for $k<2$.

\par\vskip 6ex\noindent
\mysection{Modular invariants from infinite number 
of discrete series characters}
\par\noindent
We saw that the possibility of constructing  modular invariants 
from finite number of the discrete series characters is quite limited. 
In this section we consider whether the infinite sum or the integral 
of the characters $ \chi_\mu $ leave the possibility of constructing
modular invariants.

As we saw  in section 2, modular $S$-transformation of $\chi_\mu$ 
is expressed as a superposition of infinitely many characters.
However unlike the momentum eigenstates (plane waves) in flat space, it is
not clear in what sense $\hat{sl}(2,R)$ characters $\{\chi_\mu\}_{\mu
\in R}$ are orthogonal or linearly independent. In particular, the
formula for the modular $S$-transformation of $\chi_\mu$ might not be
unique.
In fact, it may be possible to get different expressions by deforming 
the integration contours in (\ref{S1}) and (\ref{S2}).
Thus let us discuss the possibility of the modular invariants 
in a definite manner.

Here we would like to show that 
{\it the $S$-transformation of $\chi_\mu$ cannot
be expressed by using the values of $\mu $ belonging only to a finite 
interval, namely, by using  $\chi_\mu $ with $ \mu \in [ \mu_1 , \mu_2 ]$ }.
First, let us suppose that 
\eqabegin
  \chi_\mu \Bigl( \,  \frac{z}{\tau}, -\frac{1}{\tau}, 
   u + \frac{z^2}{4\tau} \, \Bigr) & = & \int_{\nu_1}^{\nu_2} d\nu 
  \ f(\nu; \mu,k) \chi_\nu (z,\tau,u)
  \comma \label{Sfinite}
\eqaend
where $ f(\nu; \mu,k) $ is some 
function which is continuous in $\nu \in [\nu_1, \nu_2] $ and independent of
$ \tau $. Using the $S$-transform of $\vartheta_1$, one finds that  
(\ref{Sfinite}) is modular invariant only if 
\eqabegin
   i \,  e^{\pi i (k-2) z^2/(2 \tau) } e^{-2\pi i \mu z/\tau }
    \qtilde^{-\frac{\mu^2}{k-2}} 
  &=&  \sqrt{-i\tau} \int_{\nu_1}^{\nu_2} d\nu \ f(\nu; \mu,k)
   \, e^{-2\pi i \nu z} q^{-\frac{\nu^2}{k-2}}
  \period  \label{qtilq}
\eqaend
This can never happen. Indeed, put $z=0$ and take the limit ${\rm
Im\;}\tau \to +\infty$.  The left hand side tends to 1 whereas the right
hand side diverges if $k>2$ and tends to 0 if $k<2$.

Similar arguments hold in the above and the following when 
the integral $\int_{\nu_1}^{\nu_2} d\nu \ f(\nu; \mu,k) $ is 
replaced with an infinite sum $ \sum_{i} \ f(\nu_i; \mu,k) $
with $\nu_i \in [ \nu_1 , \nu_2 ]$ as long as 
the sum makes sense. We will omit discussions 
in such cases.

Now let us consider a partition function of the following form:
\eqabegin
   Z(z,\tau,u) &\equiv& \int_I d \mu d \mu' \
   g(\mu,\mu') \chi_\mu (z,\tau,u) \chi^*_{\mu'} (z,\tau,u)
  \comma 
\eqaend
where $ g(\mu, \mu') $ is a weight function continuous on the 
domain $I = [\mu_1, \mu_2] \times [\mu'_1, \mu'_2] $. 
It is understood that 
the measure is zero for the discrete values of $\mu$ 
corresponding to the representations with singular vectors.  
If such a partition function is modular invariant, it follows that 
\eqabegin
\lefteqn{
\abs{ \tau } \int_I d \mu d \mu' \ g(\mu,\mu') 
  e^{-2\pi i (\mu z - \mu' z^*) } q^{-\frac{\mu^2}{k-2}} 
  (q^*)^{-\frac{\mu'^2}{k-2}} 
}\nn
\\
  &=& e^{-\pi (k-2) \Im{ \frac{z^2}{\tau} } } 
  \int_I d \mu d \mu' \ g(\mu,\mu') 
  e^{-2\pi i \bigl(\mu \frac{z}{\tau} - \mu' (\frac{z}{\tau})^* \bigr) } 
  \qtilde^{-\frac{\mu^2}{k-2}} 
  (\qtilde^*)^{-\frac{\mu'^2}{k-2}}
  \period
\label{equalityI}
\eqaend
However, since the interval $I$ is finite, a similar argument to the 
above  shows that the equality (\ref{equalityI})
cannot be true. Moreover, a similar argument holds also 
in the case where $g(\mu,\mu')$ includes distributions 
(as long as the expression makes sense). 
Therefore, we have found that 
{\it it is impossible to construct modular invariants only 
from $\chi_\mu$ with $\mu $ belonging to a finite interval 
$ \mu \in [ \mu_1 , \mu_2 ]$ even if infinitely many $\chi_\mu$ are used.} 

Thus the modular invariant partition function as a superposition of
$\chi_{\mu}\chi_{\mu'}^{*}$ is possible only if the partition
function contains arbitrarily high spin $|\mu|$.
Nevertheless, for $k>2$, the partition function becomes ill-defined 
since $L_{0}$ spectrum is not bounded from below: 
$\Delta_{j}\to -\infty$ as $ \abs{ \mu } \to \infty $.
Hence we have reached  a conclusion 
that {\it for $k>2$ it is impossible to construct modular invariants 
only from $\chi_\mu $, namely, from 
the discrete series characters without singular vectors} 
(if any procedure such as `Wick rotation' is not consistently
implemented).

Finally, we consider the partition function for $k<2$ with 
no upper bound on the spin $|\mu|$:
\eqabegin
   Z(z,\tau,u) &\equiv& \int_{-\infty}^{\infty} d \mu d \mu' \
   g(\mu,\mu') \chi_\mu (z,\tau,u) \chi^*_{\mu'} (z,\tau,u)
  \comma \label{Zint}
\eqaend
To analyze this, we introduce the Fourier transform 
\eqabegin
  g(\mu,\mu') & = & \frac{1}{(2\pi)^2} 
   \int_{-\infty}^{\infty} d\xi d\xi' \ e^{-i(\mu \xi - \mu'\xi')} 
    \hat{g} (\xi,\xi')
  \period
\eqaend
Then for $k-2<0$ one finds that  
\eqabegin
  Z \Bigl( \, \frac{z}{\tau}, -\frac{1}{\tau}, 
   u + \frac{z^2}{4\tau} \, \Bigr) &=& \frac{2}{2-k}
   \int_{-\infty}^{\infty} d \nu d\nu' \ 
   \hat{g}\bigl( \frac{4\pi \nu}{k-2}, \frac{4\pi \nu'}{k-2} \bigr)
   \, \chi_{\nu} (z,\tau,u) \chi^*_{\nu'} (z,\tau,u)
  \period
\eqaend
Thus a sufficient condition of the modular invariance is 
\eqabegin
   g(\mu,\mu') &=& \frac{2}{2-k} \, 
   \hat{g}  \bigl( \frac{4\pi \mu}{k-2}, \frac{4\pi \mu'}{k-2} \bigr)
   \period \label{ffhat}
\eqaend
It is easy to find a solution to this condition. It is
just the delta-function,
\eqabegin
   g(\mu, \mu') &=& \delta (\mu-\mu')
   \comma
\eqaend
and this gives the diagonal partition function,
\eqabegin
   Z_{\rm diag} (z,\tau,u) &=& \int_{-\infty}^{\infty} d \mu \ 
   \abs{ \chi_\mu } ^2
      \ = \ \int_{-\infty}^{0} d\mu \ \bigl( \abs{ \chi^{\rm hw}_\mu } ^2 + 
             \abs{ \chi^{\rm lw}_\mu } ^2  \bigr) \nn \\
  &=& \half e^{- 4 \pi k \Im{ u } } 
  e^{(2-k)\pi \frac{(\Im{ z } )^2}{\Im{ \tau } } }
  \sqrt{ \frac{2-k}{ \Im{ \tau } } } 
  \abs{ \, \vartheta^{-2} (z\vert \tau) \, }
  \period  \label{Zdiag}
\eqaend 
Here we have used the fact that $\chi_\mu $ with $ \mu > 0$ are
regarded as $ -\chi^{\rm lw}_\mu $ with $\mu<0$ and recovered 
the superscripts hw and lw. 
The diagonal partition function without the variable $u$ 
was discussed in \cite{PMP}. In our case, 
it is straightforward to check that $Z_{\rm diag}(z,\tau,u)$ is 
actually modular invariant because of the presence of $u$.

As pointed out also in \cite{PMP}, 
$Z_{\rm diag}(z,\tau,0)$ was discussed 
in \cite{Gawedzki} in the context of a path-integral
approach to the $SL(2,C)/SU(2)$ WZW model corresponding to Euclidean
$AdS_3$. This model has an 
$\hat{sl}_2 \times \hat{sl}_2^*$ current algebra
symmetry and the diagonal partition function may be understood also 
as the partition function of this model. However, in \cite{Gawedzki}
different spectrum seems to be  summed up. It is interesting to
consider the precise relationship between the approach here and
the one in \cite{Gawedzki}.  

\par\vskip 6ex\noindent
\mysection{Discussion}
In this paper, we discussed the modular invariants 
using the discrete series characters. The arguments hold, except for 
the last one below Eq.(\ref{Zint}),  even if we set $u=0$ and consider 
${\rm ch}_j(z,\tau,0)$ and $\chi_\mu(z,\tau,0)$.  
Our arguments indicate that the possibility of constructing 
modular invariants is very limited. 

If we use only the characters without singular vectors,
the possibility is only in the case where $k<2$ and $\chi_\mu$ with 
$ \abs{ \mu } \to \infty$ are included.
We gave such an example. 
The resulting modular invariant $Z_{\rm diag} (z,\tau,u)$ may 
be regarded as a kind of twisted partition function. 
However, its physical interpretation  is still unclear, 
in particular, regarding the factor $e^{2\pi i k u}$.

In the case in which the characters with singular vectors are allowed,
we showed that one cannot construct modular invariants from 
finite number of the characters for $V_{j}^{\rm hw}$ or $V_{j}^{\rm lw}$.
To complete the analysis, we may need to obtain the explicit forms of the 
characters with singular vectors. 

Nevertheless, it turns out that the case without singular vectors covers 
physically interesting cases and gives important implication to 
the unitarity bound (\ref{ub}). This is because the condition 
of the singular vectors (\ref{singvec}) implies that there are 
no singular vectors within (\ref{ub}).
Furthermore, since the spins $j$ in that bound belong to a finite 
interval, our results indicate  that {\it 
one cannot construct modular invariants only from the discrete 
series characters for the representations satisfying 
the unitarity bound (\ref{ub})}. This means that one cannot make 
a consistent string theory on $SL(2,R) = AdS_3$ only from the spectrum
from $V_j^{\rm hw(lw)}$ within (\ref{ub}). 

Here some comments on the relation to Ref. \cite{PMP} may be in order.
First, we discussed the cases including the representations with singular 
vectors and our analysis covered the cases of both $k>2$ and $k<2$.
In addition, we defined the character using three variables as in (\ref{chj}). 
As discussed below (\ref{chj}), this was essential to obtain a modular 
invariant (\ref{Zdiag}). Since we carried out a systematic analysis using 
the asymptotic behavior of the characters, we could derive definite 
conclusions without any ambiguities concerning which characters are linearly 
independent. Our analysis thus pinned down when the modular invariants can be 
constructed. 

Since there exist ghosts for the discrete series outside the unitarity bound,
simply adding such spectrum may not give a consistent theory.  
Therefore, the possibilities for a consistent theory seem 
(a) to use  the discrete series satisfying (\ref{ub}) 
but include some new sectors with different characters from 
$\chi_\mu$ as in \cite{HHRS}, and/or
(b) to use the spectrum of other representations as in \cite{Bars,YS}.
In any case, the string theory on 
$AdS_3$ definitely deserves further investigations. 
\vskip 10ex
\noindent
{\bf Note added}
\par\bigskip

While we were proofreading the manuscript, 
a paper \cite{MO} appeared which discusses the modular invariance
using the discrete series $\Dhw$ and $\Dlw$.
In \cite{MO}, the diagonal modular invariant (\ref{Zdiag}) is obtained
from the spectrum satisfying the (more stringent) unitarity bound 
by (i) including additional sectors along the line of \cite{HHRS}
(i.e., the possibility (a) in section 5 in our paper)
and by assuming `Wick rotation' (see the comment in section 4).
In addition, the role of $e^{2\pi i k (u-u^*)}$ in $Z_{\rm diag} (z,\tau,u)$ 
in our paper is played by the chiral anomaly term 
$e^{\pi k \frac{ ( \Im{z} )^2 }{ \Im{\tau} } }$ in \cite{MO}.
A relationship to the $SL(2,C)/SU(2)$ case \cite{Gawedzki} 
is also discussed there. 

The authors of \cite{MO} argue that the Hilbert space of the 
consistent $SL(2,R)$ string theory consists of the principal
continuous series, $\Dhw$ and $\Dlw$ including the sectors
generated by the Weyl reflections.
This is consistent with the discussion in section 5.  
%
\newpage
\csectionast{Acknowledgements}
We would like to thank I. Bars, J. de Boer, A. Giveon, N. Ishibashi
and S.-K. Yang for discussions and P.M. Petropoulos 
for useful correspondences. We would also like to thank the 
organizers of Summer Institute '99 held at Fuji-Yoshida, Japan, 
7-21 August, 1999, where this work was started.
The work of A.K. is supported in part by the Grant-in-Aid 
for Scientific Research (No. 11740140) from the Ministry of Education, 
Science, Sports and Culture of Japan.
%
%
%
\def\thebibliography#1{\list
 {[\arabic{enumi}]}{\settowidth\labelwidth{[#1]}\leftmargin\labelwidth
  \advance\leftmargin\labelsep
  \usecounter{enumi}}
  \def\newblock{\hskip .11em plus .33em minus .07em}
  \sloppy\clubpenalty4000\widowpenalty4000
  \sfcode`\.=1000\relax}
 \let\endthebibliography=\endlist
\vskip 10ex
\csectionast{References}

\end{document}